\documentclass[final,narrowdisplay]{elsart5p}

\usepackage{amsmath}
\usepackage{amssymb}
\usepackage{graphicx}

\begin{document}

\begin{frontmatter}
\title{The control of iron oxidation state during FeO and olivine crystal growth}
\author{D. Klimm and S. Ganschow}
\address{Institute of Crystal Growth, Max-Born-Str. 2, 12489 Berlin, Germany}
\date{}

\begin{abstract}
Crystal growth experiments (micro-pulling down or Czochralski methods, respectively) and DTA/TG measurements with Fe$^{2+}$ containing olivines (fayalite--forsteri\-te solid solution) and with FeO (wustite) are performed. For both substances the oxygen partial pressure $p_{\text{O}_2}$ of the growth atmosphere had to be adjusted within the stability region of Fe$^{2+}$ for all temperatures ranging from room temperature to the melting point. The formation of Fe$^{3+}$ (Fe$_3$O$_4$, Fe$_2$O$_3$) had to be avoided. The adjustment of $p_{\text{O}_2}$ could be obtained by a mixture of argon, carbon dioxide and carbon monoxide. Thermodynamic equilibrium calculations show, that mixtures of an inert gas (e.g. argon) with another gas or gas mixture that supplies oxygen at elevated temperature (e.g. CO$_2$/CO) are superior to the use of inert gases with constant oxygen admixture. The reason is that the Ar/CO$_2$/CO mixture adjusts its oxygen concentration with temperature in a way similar to that needed for the stabilization of Fe$^{2+}$.
\end{abstract}

\begin{keyword}
A1 Phase equilibria \sep A2 Czochralski method \sep B1 Oxides

\PACS 61.50.Nw \sep 77.84.Bw \sep 81.10.-h \sep 81.30.-t
\end{keyword}

\end{frontmatter}
\thispagestyle{empty}

\section{Introduction: Thermodynamic background}

\begin{figure}[htbp]
\centering
\includegraphics[width=0.48\textwidth]{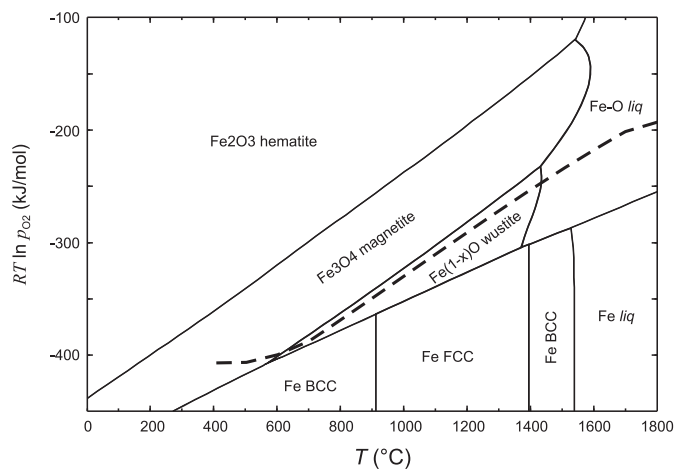}
\caption{Solid lines: Predominance (Ellingham) diagram $RT\ln p_{\text{O}_2}$ vs. $T$ for the Fe--O system at a total pressure $p=1$~bar. Dashed lines: $RT\ln p_{\text{O}_2}$ vs. $T$ for a gas mixture 85\% Ar/10\% CO$_2$/5\% CO (cf. later in the text) calculated with FactSage \cite{FactSage5_2}.}
\label{fig:Ellingham}
\end{figure}

Depending on temperature $T$, equilibrium oxygen partial pressure $p_{\text{O}_2}$ and chemical environment different valencies are stable in the iron--oxygen system: +3 (hematite Fe$_2$O$_3$; magnetite Fe$_3$O$_4$), +2 (magnetite; wustite FeO), and 0 (metallic Fe). In principle it would be possible to draw a $x$ vs. $T$ phase diagram for the binary system $(1-x)$ O --- $x$ Fe, but from such composition--temperature diagram information on the $p_{\text{O}_2}$ that is necessary to stabilize a desired condensed phase cannot be obtained. Predominance diagrams in the $T-RT\ln p_{\text{O}_2}$ plane (Fig. \ref{fig:Ellingham}) are an alternative way to plot stability regions of different condensed phases \cite{Pelton91}. If all condensed phases have fixed stoichiometry and negligible vapor pressure, the predominance regions of different phases are separated by straight lines. In practice, for all iron oxides deviations from stoichiometry are reported.

One distinct oxide can only be kept thermodynamic stable, if $p_{\text{O}_2}$ is adjusted to an appropriate level within its predominance region. The problem of valency stabilization of iron (in this case for the spinel MgFe$^{3+}_2$O$_4$) was solved recently \cite{Turkin04} by heating stoichiometric mixtures in air filled sealed ampoules to $1100^{\:\circ}$C. Under such conditions $p_{\text{O}_2}$ within the ampoule reaches $\approx1$~bar and limits oxygen loss. Fig. \ref{fig:Ellingham} shows, that for $T<580^{\:\circ}$C Fe$_3$O$_4$ and hence Fe$^{3+}$ is in direct equilibrium with metallic Fe, as wustite its unstable. Experimental evidence for equilibrium between Fe$^{3+}$ and Fe$^0$ under the conditions of the earth's mantle was given recently \cite{Frost04}.

\begin{figure}[htbp]
\centering
\includegraphics[width=0.48\textwidth]{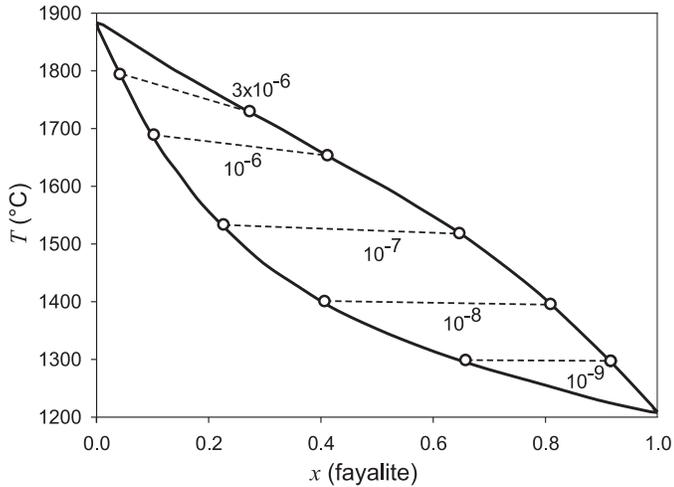}
\caption{Quasi-binary phase diagram forsterite--fayalite $(1-x)$ Mg$_2$SiO$_4$ + $x$ Fe$_2$SiO$_4$. Parameters: Oxygen partial pressure at the \emph{solidus} and \emph{liquidus} \cite{FactSage5_2}.}
\label{fig:Olivine-PD}
\end{figure}

If the predominance region of a phase is narrow, it is sometimes impossible to find one constant level of $p_{\text{O}_2}$ that stabilizes this phase over an extended $T$ region. On the other hand, crystal growth processes are performed within temperature gradients that span usually hundreds of Kelvins. In such cases, the inevitable adjustment $p_{\text{O}_2}(T)$ can often be obtained by mixtures of gases that contain at least one oxygen supplying component. Thermodynamic equilibrium calculations allow to determine such compositions, for which $p_{\text{O}_2}(T)$ is on a level that stabilizes the desired oxide over an extended $T$ region. This paper will show, that a mixture of 85\% Ar, 10\% CO$_2$ and 5\% CO stabilizes Fe$^{2+}$ in oxidic phases as olivines (Mg,Fe)$_2$SiO$_4$ (melting point $T_\text{f}=1216\ldots1881^{\:\circ}$C depending on composition, Fig. \ref{fig:Olivine-PD}) and wustite Fe$_{1-x}$O ($T_\text{f}\approx 1400^{\:\circ}$C). In anticipation of this result the calculated dependence $RT\ln p_{\text{O}_2}(T)$ of this gas mixture is plotted in Fig. \ref{fig:Ellingham} as dashed line crossing the wustite (Fe$^{2+}$) predominance region.

\section{Experimental}

A NETZSCH STA 409C with standard SiC heater furnace and heating/cooling rates of $\pm10$~K/min was used for thermal analysis. Samples were placed in Pt95\%-Au5\% DTA crucibles without lid in combination with a sample holder with thermocouples type "S" (Pt90\%Ph10\%--Pt). The DTA sample holder was calibrated for temperature and (to get at least semiquantitative DSC results) for sensitivity at the melting points of Zn, Au and Ni and at the phase transformation points of BaCO$_3$. As usual \cite{Hohne96}, the extrapolated peak onset temperatures are used as melting points in the following ($T_\text{f}=T_\text{e}$).

\begin{figure}[htbp]
\centering
\includegraphics[width=0.40\textwidth]{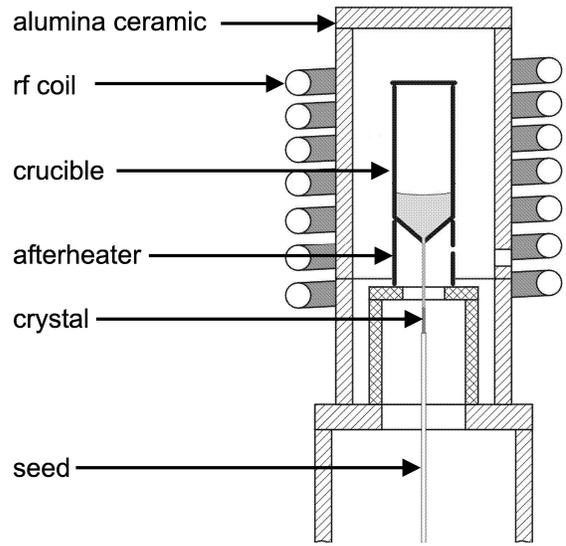}
\caption{Setup for $\mu$-pulling down. Iridium crucible design: diameter 15~mm, height 35~mm, capillary aperture diameter 0.7~mm, length of capillary 1.3~mm.}
\label{fig:Micro-PD}
\end{figure}

Crystal growth was performed with the micro-pulling-down ($\mu$-PD) method (Fig. \ref{fig:Micro-PD}). Fibers were pulled from iridium crucibles with rates of $\approx0.5$~mm/min. Iron(II) oxalate dihydrate Fe(COO)$_2\cdot$2H$_2$O with 5N (99.999\%) purity was used as starting material for the growth of wustite crystals. Pieces of a Czochralski grown forsterite (Mg$_2$SiO$_4$) single crystal previously obtained from MgO and SiO$_2$ (5N) powders were used as starting material for the $\mu$-PD of forsterite. Mixtures of Mg$_2$SiO$_4$ and of iron oxalate + silicon dioxide in the stoichiometric molar ratio Fe(COO)$_2\cdot$2H$_2$O : SiO$_2$ = 2 : 1 were used for the growth of olivine mixed crystals $(1-x)$ Mg$_2$SiO$_4$ + $x$ Fe$_2$SiO$_4$ with $0\leq x\leq 0.2$. A piece of Czochralski grown $[100]$ Mg$_2$SiO$_4$ was used as seed for $\mu$-PD.

\section{Results and Discussion}

\subsection{Thermal analysis of wustite}

\begin{figure}[htbp]
\centering
\includegraphics[width=0.40\textwidth]{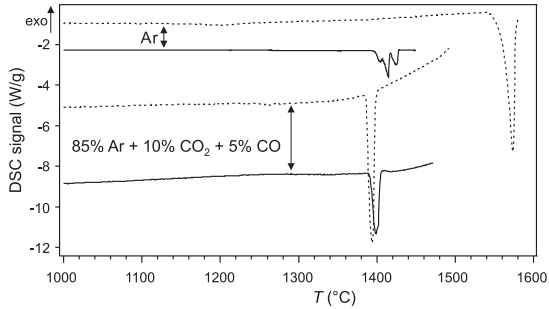}
\caption{$1^\text{st}$ melting (solid lines) and $4^\text{rd}$ melting (dashed lines) of iron oxide obtained by in situ thermolysis of Fe(COO)$_2\cdot$2H$_2$O in 5N argon or in the designated gas mixture.}
\label{fig:DSC-Gas-Ar}
\end{figure}

Fe(COO)$_2\cdot$2H$_2$O contains iron as Fe$^{2+}$. It was reported recently \cite{Bertram04}, that the thermolysis
\begin{eqnarray}
\lefteqn {\text{Fe(COO)}_2\cdot\text{2H}_2\text{O} \stackrel{\textrm{argon}}{\longrightarrow}} \nonumber \\ &&\hspace{1.5cm}\text{FeO + CO}_2 \uparrow \text{+ CO} \uparrow \text{+ 2 H}_2\text{O} \uparrow \label{eq:Fe-Ox-1} \\
\lefteqn {\text{2 Fe(COO)}_2\cdot\text{2H}_2\text{O} \stackrel{\textrm{air}}{\longrightarrow}} \nonumber \\ && \hspace{1.5cm}\text{Fe}_2\text{O}_3 + \text{3 CO}_2 \uparrow + \text{ CO} \uparrow + \text{ 4 H}_2\text{O} \uparrow \label{eq:Fe-Ox-2}
\end{eqnarray}
leads under "inert" (5N argon) or oxidative (air) conditions to a theoretical mass loss of 60.06\% (\ref{eq:Fe-Ox-1}) or 55.63\% (\ref{eq:Fe-Ox-2}), respectively. Independent on atmosphere, \linebreak Fe(COO)$_2\cdot2$H$_2$O decomposes in a 2 step process loosing initially water at $\approx200^{\:\circ}$C and finally CO and CO$_2$ around $400^{\:\circ}$C (Fig. 5 in \cite{Bertram04}). Fig. \ref{fig:DSC-Gas-Ar} in the present paper shows only the DSC melting peaks of the oxalate after complete decomposition to iron oxide. The melting temperature in 5N argon ($T_\text{f}=1398.2^{\:\circ}$C) is almost identical with the theoretical value that has to be expected for Fe$_{1-x}$O ($T^\text{FeO}_\text{f}\approx1400^{\:\circ}$C depending on $x$, Fig. \ref{fig:Ellingham}).

Position and shape of the melting peak in argon show bad reproducibility; in Fig. \ref{fig:DSC-Gas-Ar} the peak is heavily distorted. If the measurement is repeated several times under identical conditions, the peak shifts gradually to higher $T$. In the $2^\text{nd}$ run (not shown in Fig. \ref{fig:DSC-Gas-Ar}) $T_\text{f}=1426^{\:\circ}$C was found for the same sample. After several DSC heating/cooling runs the peak remains stable at $T_\text{f}\approx1562^{\:\circ}$C (upper dashed line in Fig. \ref{fig:DSC-Gas-Ar}). This temperature corresponds to the melting point of magnetite (Fig. \ref{fig:Ellingham}).

These observations can be explained by a gradual oxidation of wustite to magnetite in the nominally "inert" argon atmosphere -- practically the term "inert" is inadequate. As the gas has a purity of 99.999\% it contains up to 0.001\% ($10^{-5}$) unspecified impurities. Only if these impurities were completely free of oxygen, $p_{\text{O}_2}\approx0$ would hold for the atmosphere. (Actually a lower limit is determined by the oxygen partial pressure of constructive parts made of alumina. This value is for O atoms $p_{\text{O}}=6\times10^{-12}$~bar and for O$_2$ molecules $p_{\text{O}_2}=3\times10^{-14}$~bar at $1400^{\:\circ}$C \cite{FactSage5_2}.) If one assumes air with 21\% O$_2$ as impurity, $p_{\text{O}_2}\lessapprox2\times10^{-6}$~bar results as approximate value for the "inert" argon atmosphere used for the DSC measurements and $RT\ln p_{\text{O}_2}\approx-180$~kJ/mol can be calculated at $1400^{\:\circ}$C = 1673~K. Almost identical values are calculated if 5N N$_2$ is used instead of 5N Ar. Fig. \ref{fig:Ellingham} shows, that the point [$1400^{\:\circ}$C, $-180$~kJ/mol] is far within the predominance region of magnetite. The phase boundary to wustite is crossed not before two orders of magnitude lower oxygen partial pressures  at $p_{\text{O}_2}\lessapprox3\times10^{-8}$~bar. From this result it follows, that it is impractible to maintain molten wustite in argon atmosphere without gradual oxidation to magnetite.

As mentioned in the introduction, the application of specific gas mixtures that can buffer $p_{\text{O}_2}(T)$ is an alternative way to maintain Fe$^{2+}$ stable. Mixtures of carbon dioxide and carbon monoxide are well suited for this purpose, as the equilibrium reactions

\begin{eqnarray}
\text{CO}_2  & \rightleftarrows & \text{CO} + \frac{1}{2} \text{O}_2 \label{eq:CO2} \\
2 \text{CO}  & \rightleftarrows & \text{CO}_2 + \text{C} \label{eq:CO}
\end{eqnarray}

stabilize $p_{\text{O}_2}$ for a given $T$ to a large extend. Additional Ar dilutes the hazardous CO and does not influence the equilibria (\ref{eq:CO2}) and (\ref{eq:CO}) substantially, if the CO and CO$_2$ concentrations remain in the order of several per cent. For a mixture 85~mol-\% Ar + 10~mol-\% CO$_2$ + 5~mol-\% CO one calculates at $1400^{\:\circ}$C $p_{\text{O}_2}=9.65\times10^{-9}$~bar. This partial pressure is not very sensitive against additional O$_2$ impurities: If as much as 0.1\% (1000~ppm) O$_2$ are added, one obtains $p_{\text{O}_2}=1.09\times10^{-8}$~bar for the gas mixture. The same addition of O$_2$ to ideally pure Ar would result in $p_{\text{O}_2}=10^{-3}$~bar!

The two lower curves in Fig. \ref{fig:DSC-Gas-Ar} show iron oxide melting peaks from the first and the fourth DSC heating runs of a Fe(COO)$_2\cdot$2H$_2$O sample in the gas mixture. The measured values are $T_\text{f}=1386.5^{\:\circ}$C and $T_\text{f}=1385.0^{\:\circ}$C, respectively. These almost identical values give evidence, that Fe$^{2+}$ could be stabilized with the Ar/CO$_2$/CO atmosphere in the investigated region $T\leq1480^{\:\circ}$C. For $T\gg1500^{\:\circ}$C the evaporation of wustite under partial decomposition to Fe and O was found to be remarkable. The metallic iron alloys and destroys the DTA thermocouples making it impossible to extend the thermoanalytic measurements considerable above the $T$ range shown in Fig. \ref{fig:DSC-Gas-Ar}. This restriction does not apply after several melting runs in 5N Ar that lead to the formation of magnetite, as $p_{\text{Fe}}$ is more than one order of magnitude smaller for Fe$_3$O$_4$ as compared to FeO \cite{FactSage5_2} -- therefore the melting of magnetite at $T_\text{f}\approx1562^{\:\circ}$C (upper dashed line in Fig. \ref{fig:DSC-Gas-Ar}) could be measured.

\subsection{Crystal growth of wustite and olivine}

Wustite was grown only within the Ar 85\% + CO$_2$ 10\% + CO 5\% gas mixture. Unfortunately it turned out, that iridium is wetted by the wustite melt very well. Accordingly, it was not possible to control the diameter of the growing crystal. However at very high pulling rates (20\ldots50~mm/min) the diameter self-adjusted to $\approx5$~mm, probably due to insufficient material transport through the capillary. The largest single crystalline region within this body had a diameter of 3~mm. X-ray powder diffraction analysis confirmed, that the black body consists of pure wustite.

\begin{figure}[htbp]
\centering
\includegraphics[width=0.40\textwidth]{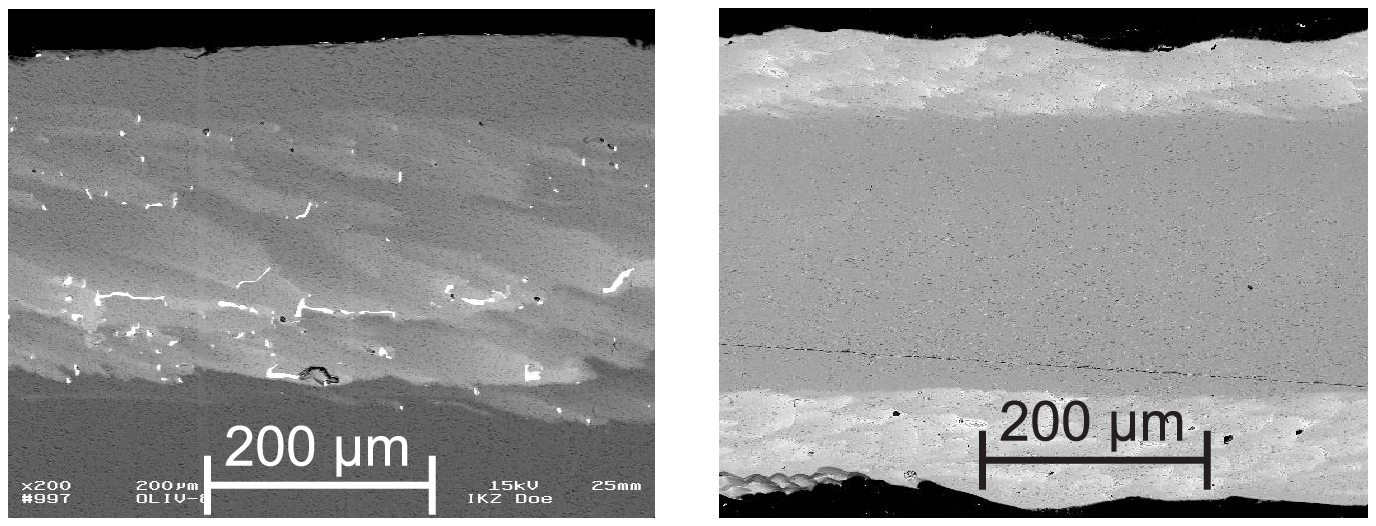}
\caption{REM photographs (BSE contrast) of $(1-x)$ Mg$_2$SiO$_4$ + $x$ Fe$_2$SiO$_4$ mixed crystals with identical fayalite concentration $x=0.111$.
\newline Left: grown in 5N N$_2$. Right: grown in Ar/CO$_2$/CO mixture.}
\label{fig:BSE}
\end{figure}

The growth of olivine mixed crystals $(1-x)$ Mg$_2$SiO$_4$ + $x$ Fe$_2$SiO$_4$ with $0\leq x\leq 0.2$ was performed in 5N nitrogen as well as in the Ar 85\% + CO$_2$ 10\% + CO 5\% gas mixture. As expected from the thermoanalytic measurements and from the thermodynamic equilibrium calculations reported above, the crystals that were grown in nitrogen contained Fe$^{3+}$. EDX analysis revealed that the bright inclusions in Fig. \ref{fig:BSE} (left) are Fe$_3$O$_4$. Almost no inclusions could be found in olivines that were grown in the gas mixture. The concentrations of magnesium and iron were found to be constant along the crystal fiber length of up to 100~mm as well as across the main part of the cross section with except of an outer rim with $\approx200~\mu$m thickness. In this rim $x_\text{Fe}/(x_\text{Fe}+x_\text{Mg})$ is about 4 times larger than in the center. The rim that is visible as bright region in the right Fig. \ref{fig:BSE} results from peculiarities of the flow regime of the melt after passing the capillary and related segregation phenomena and will not be discussed here in detail. From the phase diagram in Fig. \ref{fig:Olivine-PD} follows that the first crystallized parts in the center of the crystal are Fe depleted. Fe$_3$O$_4$ in the still liquid rim is formed by oxidation from the atmosphere. Unfortunately, details of the convection regime in this rim are unknown but Fig. \ref{fig:BSE} (left) suggests, that the strong convection is responsible for the enrichment of the Fe$_3$O$_4$ particles near the rim/core interface.

The Fe$_3$O$_4$ inclusion formation in N$_2$ grown olivines results from the too high oxygen partial pressure being constant during the whole growth process. $p_{\text{O}_2}\approx2\times10^{-6}$~bar was calculated above as typical value of 5N "inert" gases as Ar or N$_2$. This value corresponds approximately to the equilibrium oxygen partial pressure of Mg rich material with $x\leq0.2$ Fe$_2$SiO$_4$ (Fig. \ref{fig:Olivine-PD}). However, the equilibrium oxygen partial pressure over the Fe enriched material in the rim is considerably lower. Therefore oxidation
\begin{equation}
2\;\text{Fe}^{2+} + \frac{1}{2}\;\text{O}_2 \rightarrow 2\;\text{Fe}^{3+} + \text{O}^{2-}
\end{equation}
by the too high $p_{\text{O}_2}$ takes place especially there.

The oxidation can be avoided, if the growth is performed in the gas mixture. Fig. \ref{fig:Ellingham} shows, that the special composition of 85\% Ar + 10\% CO$_2$ + 5\% CO adjusts over an extended $T$ range of over 1000~K $p_{\text{O}_2}$ to levels within the predominance region of Fe$^{2+}$. (Obviously the Gibbs free energy change of Fe$^{2+}$ in the olivine solid and liquid phases as compared to pure Fe$_{1-x}$O is so small that it does not influence the predominance regions too much.)

\section{Summary}

Thermodynamic equilibrium calculations, thermal analysis and $\mu$-PD crystal growth experiments were performed for wustite and for forsterite-fayalite solid solutions. It was shown to be inevitable, that an adjustment of the oxygen partial pressure in the atmosphere $p_{\text{O}_2}(T)$ is performed over several orders of magnitude if $T$ changes during the crystal growth process by several 100~K. Undesired valency changes result, if the $p_{\text{O}_2}(T)$ adjustment fails. If Ar or N$_2$ with 5N purity as "inert" gases are used as atmosphere, Fe$^{2+}$ could not be stabilized; instead partial oxidation to Fe$^{3+}$ lead to the formation of Fe$_3$O$_4$. Using an atmosphere of 85\% Ar + 10\% CO$_2$ + 5\% CO, almost perfect adjustment of $p_{\text{O}_2}(T)$ could be obtained for all $T\lessapprox1900^{\:\circ}$C.

\begin{ack}

The authors express their gratitude to J. Doerschel for performing TEM and EDX measurements. V. Alex is thanked for X-ray phase analysis. R. Sommer (Kistler Instruments AG Winterthur, Switzerland) is acknowledged for his valuable contribution.

\end{ack}

\end{document}